\documentclass{ws-ijmpa}
\usepackage[super,compress]{cite}
\usepackage{graphicx}
\usepackage{xspace}
\usepackage{todonotes}
\usepackage{hyperref}
\usepackage{cleveref}
\usepackage[binary-units=true]{siunitx}

\begin{document}
\markboth{Nils Braun, Thomas Kuhr}{Software for Online Reconstruction and Filtering at the Belle~II Experiment}

%
\catchline{}{}{}{}{}
%

\title{Software for Online Reconstruction and Filtering at the Belle~II Experiment}

\author{Nils Braun}

\address{Karlsruhe Institute of Technology\\
nilslennartbraun@gmail.com}

\author{Thomas Kuhr}

\address{Ludwig-Maximilians-Universit\"at M\"unchen\\
Thomas.Kuhr@lmu.de}

\maketitle

\begin{history}
\received{Day Month Year}
\revised{Day Month Year}
\end{history}

\begin{abstract}
The Belle II experiment is designed to collect 50 times more data than its predecessor.
For a smooth collection of high-quality data, a robust and automated data transport and processing pipeline has been established.
We describe the basic software components employed by the high level trigger.
It performs a reconstruction of all events using the same algorithms as offline, classifies the events according to physics criteria, and provides monitoring information.
The improved system described in this paper has been deployed successfully since 2019.
\keywords{High Level Trigger; Belle II; Software.}
\end{abstract}

\ccode{PACS numbers:}


\newcommand{\zmq}{\O{}MQ\xspace}

\section{Introduction}
The Belle II experiment~\cite{Abe:2010gxa} at the SuperKEKB $e^+e^-$ accelerator~\cite{Akai:2018mbz} is a next-generation B factory experiment~\cite{Bevan:2014iga} with a rich physics program~\cite{Kou:2018nap}.
A particular focus is on searches for physics beyond the standard model by high-precision measurements of CP asymmetries and rare decays of B mesons, charm hadrons, and tau leptons.
Further objectives are the improved measurements of quark mixing matrix elements~\cite{Kobayashi:1973fv}, which are fundamental parameters of the standard model, and studies of (new) hadronic states.

To fulfill its mission, the Belle II experiment will collect a significantly larger data set than its predecessor, Belle~\cite{Abashian:2000cg}.
The goal of a 50 times larger data set is tackled by reducing the beam spot size and increasing the beam currents.
The design luminosity of SuperKEKB is a factor 40 higher than what was achieved at $e^+e^-$ colliders so far.
The higher instantaneous luminosity in the nano-beam scheme also leads to more background that is mitigated by upgrades of the detector.

The innermost part of the Belle II detector is a pixel vertex detector (PXD) based on the DEPFET technology~\cite{Kemmer:1986vh} with two cylindrical layers at radii of 14~mm and 22~mm.
It is surrounded by four layers of double-sided silicon strip sensors, the silicon vertex detector (SVD).
The main device for the measurement of the momentum of charged particles is the central drift chamber (CDC), a gaseous detector with an outer radius of 113~cm.
In addition to specific energy loss in the CDC, particle identification information is provided by Cherenkov radiation in the time of propagation (TOP) and aerogel ring-imaging Cherenkov (ARICH) detectors in the barrel and forward regions, respectively.
An electromagnetic calorimeter (ECL) detects photons and identifies electrons.
A solenoid provides a magnetic field of 1.5~T in the enclosed detectors.
The flux return iron yoke is instrumented to detect long-lived neutral kaons and identify muons (KLM).

An online selection of events is done in two stages: at a level 1 (L1) trigger implemented in hardware and a High Level Trigger (HLT) implemented in software.
The main challenge is to suppress (radiative) Bhabha events while maintaining near 100\% efficiency for $Y(4S) \to B\bar{B}$ events, hadron production from continuum events, and tau pair events.
In L1, (two-dimensional) track segments in the CDC, energy measurements in 4x4 cells in the ECL, timing information from the TOP, and a matching of CDC tracks to KLM signals for muon identification are available.
The expected L1 output rate is 20~kHz at design luminosity.

If L1 accepts an event, the detector signals are read out via a unified link to common boards in readout PCs (COPPER)\cite{Yamada2015}.
After passing an event-building network, the HLT runs the same reconstruction algorithms that are use for the offline processing.
These include a determination of the time of the $e^+e^-$ collision, the finding and fitting of tracks in CDC and SVD, the clustering of energy deposits in the ECL, an extrapolation of tracks to the outer detectors, and the calculation of particle identification likelihoods based on dE/dx and matched signals in TOP, ARICH, ECL, and KLM.
Coarse calibration constants are stored in the conditions database and are provided via a network file system.
Based on the reconstructed objects, the events are categorized and selected.
The criteria are also stored in the conditions database so that it is easy to update and keep track of them.
A further important task of the HLT is to create histograms for data quality monitoring (DQM).
The DQM histograms and an event display allow the shifters to detect in real time issues in the detector operation and data taking.

A special PXD readout system is implemented because, with its 8 million channels, it yields a data rate that is one order of magnitude higher than that of the other detectors combined.
With a readout time of 20~$\mu$s, the PXD integrates over about 5000 bunch crossings that occur at a frequency of 250~MHz.
To reduce the data rate to a level that can be handled by the data acquisition system, regions of interest (ROIs) are determined by the HLT based on the extrapolation of tracks reconstructed in the CDC and SVD.
If the ROI selection is enabled, only clusters within the ROIs are passed on to a second stage event builder.
The events with PXD information are written to storage from where they are transferred to the KEK computing center for permanent archival.
At design luminosity, the expected output rate is 10~kHz and the expected raw data event size about 200~kB, where half is for PXD information.
A subset of events is processed in an express reconstruction to enable rapid monitoring of the PXD data quality.

In the 2018 commissioning phase, called phase 2, backgrounds were studied and collision data were recorded with only one module per PXD and SVD layer.
After the installation of the PXD and SVD, the physics run, called phase 3, started in March 2019.

\section{Online Reconstruction and Trigger Software}
\subsection{Belle~II Analysis Software Framework}
The same framework, the Belle~II Analysis Software Framework, basf2~\cite{Kuhr:2018lps}, is used on the HLT, the express reconstruction, the offline processing, the Monte-Carlo production, and the physics analysis.
It is written in C++ and has a python interface for configuration.
The configuration defines a sequence of dynamically loaded modules and sets their parameters.
Events are processed according by all modules in this so-called path.
A module can have a return value that can redirect the processing of the event to an alternative path.
This feature is used for the implementation of the event filtering by the HLT.

Event data are shared among modules via a globally accessible DataStore.
The ROOT~\cite{Brun:1997pa} data format is used for persistency.
Access to event and conditions data is provided by the framework in a transparent way via smart-pointer-like objects.

On the HLT, several variables, identified by strings, are calculated and then used to define physics-process inspired event categories.
The selection criteria on the values of these variables, with optional prescale factors, are stored in the conditions database.
Several selections are combined for the trigger menu.
The storage of selection criteria and trigger menus in the conditions database allows their adjustment according to data taking conditions and physics priorities without the need to modify the code.
It also provides a bookkeeping of the high level trigger conditions under which the data was taken.
The result of each single selection can be to either accept the event, to reject it, or to take no decision.
A final binary event decision takes all single selections into account.
The event-by-event trigger decisions are available to analysis jobs via the DataStore.

\subsection{Trigger Menu}
The main physics background consists of Bhabha-scattering events.
They are identified in the HLT by two high-energy clusters in the ECL that are approximately back-to-back in the center of mass system.
Alternatively, one cluster could be replaced by a track in the Bhabha selection criterion.
The Bhabha triggers are typically used as vetos in the physics trigger definitions.

Hadronic events are identified with high efficiency by requiring at least three tracks.
An alternative selection of these events is implemented by requiring three clusters in the ECL with some minimal and maximal energy.
The latter, in particular, selects tau-pair events.
Further low-multiplicity events are selected by trigger definitions based on two tracks or single clusters with tuned momentum or energy and polar angle requirements.

\subsection{Reconstruction Algorithms}
The HLT decision is based on the reconstructed objects.
In the first step, a local reconstruction in each sub-detector is performed.
This is typically a clustering of signals assumed to originate from a common parent particle.
In the ECL, the algorithm reconstructs clusters under different hypotheses for electromagnetically and hadronically interacting particles, as they have different shower shapes.

The reconstruction of tracks begins with the finding of candidates in the CDC with a global and local approach.
After merging the results of both algorithms, hits in the SVD are attached using a combinatorial Kalman filter.
Low momentum tracks that are not reconstructed in the CDC are addressed by a cellular automaton based track finder in the SVD with an optimized reduction of considered hit combinations.
Finally, the track parameters are determined with GenFit2~\cite{Bilka:2019ang}.

The tracks are extrapolated to the outer detectors and clusters in the ECL and KLM are matched to the tracks.
This information and the assignment of signals in the TOP and ARICH detectors to tracks are used to calculate per-track particle identification likelihoods for electrons, muons, pions, kaons, protons, and deuterons.

\subsection{Machine Learning in the Online Reconstruction}
Several steps in the reconstruction require decisions that take into account multiple discriminating variables.
Multivariate analysis (MVA) and, in particular, machine learning techniques are well suited to address these challenges.
The basf2 framework includes a package for multivariate analysis.
It provides an interface to various machine learning algorithms and uses the conditions database for the storage of the trained models.
At Belle II, the FastBDT (fast boosted decision tree) algorithm~\cite{Keck:2017gsv} is often chosen because of its optimized inference execution time.

The ECL reconstruction uses machine learning for the analysis of the spatial shower shape and of the pulse shape time signal.
In the tracking, BDTs are employed for the assessment of the quality of hits, track segments, and combinations of track candidates.
The identification of neutral long-lived kaons in the KLM relies on machine learning, too.
A further application of machine learning that is under development is the reconstruction of clusters in the SVD.
And, of course, machine learning plays an important role in many analyses and analysis tools that are applied after the reconstruction.

\section{HLT Data Transport Implementation}
The HLT is one of the key components in the Belle~II data acquisition.
It executes the reconstruction and the event filtering described in the previous section and assures acceptable data quality by monitoring important physics quantities in the DQM as well as displaying selected events to the shifters.
Additionally, it is responsible for transporting the data from the detector event builder to the storage system.
This section describes the data transport system used in the HLT with an emphasis on the new implementation based on the open-source library \zmq~\cite{zeromq}, which was deployed in fall 2019.

\subsection{Data flow and Components}
For leveraging the inherent parallelism due to the independence of the collision events, the HLT computing farm consists of multiple independent units.
The final setup will contain around 20 HLT units~\cite{Lee2011}, which include 12-20 worker nodes each doing the reconstruction and event filtering on each of their CPUs.
Additionally, each unit contains a dedicated input, output and storage node.
For the nominal instantaneous luminosity, the expected input event rate to the full HLT system is \SI{20}{kHz} with a data size around \SI{100}{\kilo\byte} for each event without PXD data.
Input nodes have a fiber network connection to the detector readout system.
The event builder processes running on these machines collect the detector data for one collision event from each subsystem readout and merges them into a single event package, which can be retrieved by the HLT framework via TCP.
The data for a single event are distributed to one of the processes running on a CPU of a worker node, which decides whether to retain the event after performing a full event reconstruction.
The binary detector data of all events (or only the meta-data in case of a dismissed event) is collected at the output node.
Information on the selected ROIs is sent to the PXD readout system whereas the event data travels on to the HLT storage nodes.
These nodes store the data together with the PXD information in the ROIs until it is retrieved from the final storage at the KEK computing center~\cite{Asner2013}.


\subsection{HLT Data Transport Framework During Phase 2}
In Ref.~\cite{Lee2011}, the framework for performing the data and quality information transportation utilized in the data taking of Phase 2 is described.
The ideas of the implementation follow the framework used in the Belle experiment and can also be utilized for performing multiprocessing calculations during development or offline reconstruction.

\begin{figure}
    \centering
    \includegraphics[width=\linewidth]{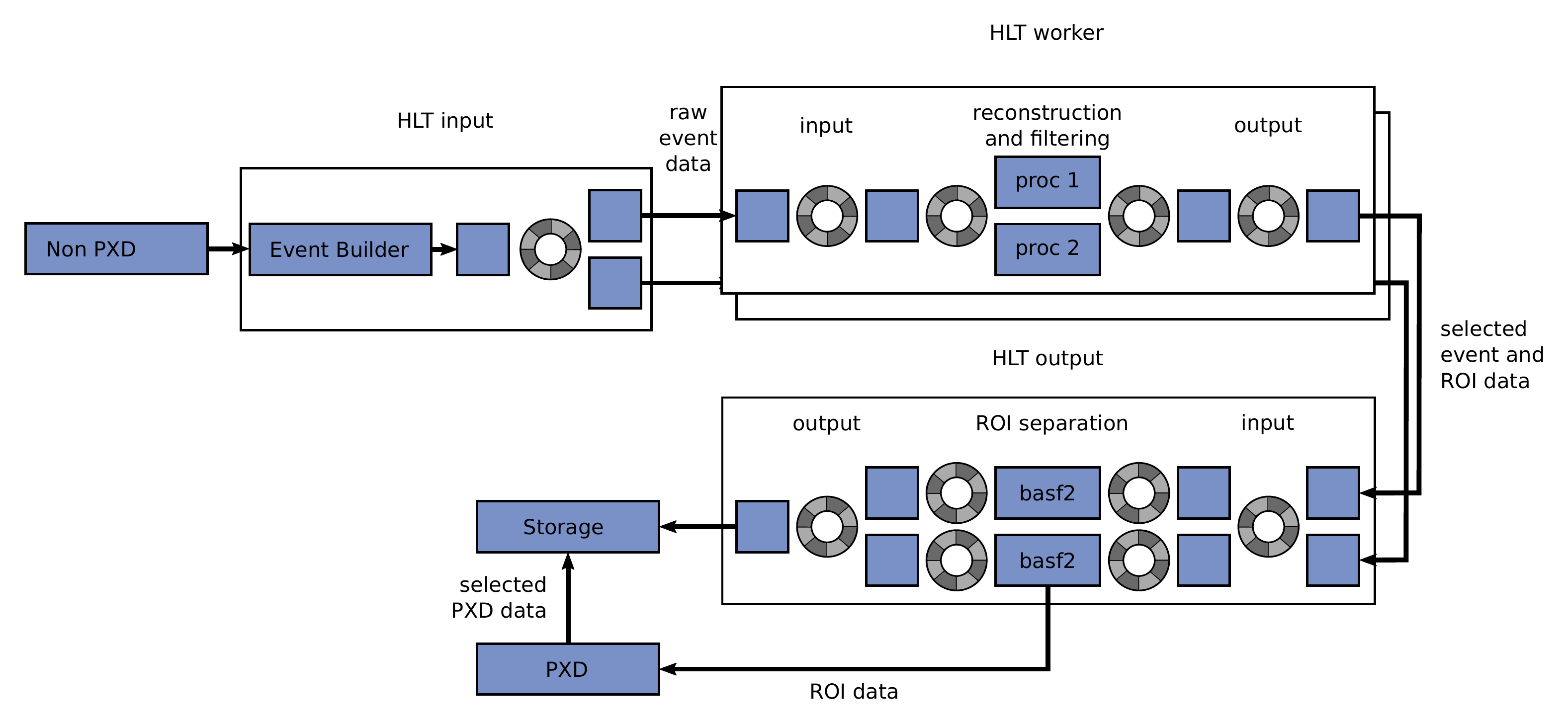}
    \caption{
        Data Transport Implementation on the HLT of Belle~II used during the data taking of Phase~2 utilizing shared ring buffers.
        Only a single worker with two processes is shown and the DQM histogram transport is omitted for better visibility.
        After Ref.~\cite{Lee2011}.
    }
    \label{fig:hlt-rb}
\end{figure}

The core component is a custom shared-memory ring buffer implementation.
Multiple processes can read and write to the buffer.
Consistency is controlled via System V semaphores to serialize parallel accesses.
Ring buffers are utilized between two subsequent stages of the data flow on the same machine for decoupling and to interchange data.
The communication between nodes is handled via a custom TCP protocol.
\Cref{fig:hlt-rb} shows the data transportation system of the HLT.

\subsection{Needs for a new HLT Data Transport Software Implementation}


The data transport software described in the previous section was used until fall 2019.
After several improvements, a stable operation was achieved serving peak data rates up to \SI{7}{\kilo\hertz} with 5 HLT units.
However, many open issues could not be solved by incremental small changes to the software.
The semaphore-based ring buffers require a persistent state, which is not coupled to a specific process (as it is used for inter-process communication).
After an abort of the data taking -- especially but not only in case of unforeseen incidents -- the residual state is indeterminate and might be invalid.
The recovery requires complex and error-prone procedures, which too often imply manual expert intervention.
Additionally, the rigid structure of the data transport makes it impossible to establish additional, non-data communication.
This non-data communication could however be used beneficially to initialize the reconstruction process even before the first event is processed and thus reduce the startup time or to properly wait until all events are processed.
Both are only possible by waiting a predefined fixed amount of time in the old schema.
In some cases, this might lead to delays of up to \SI{5}{\minute} between the transitions of two runs.
Last but not least, the custom Belle~II implementations of the ring buffer and the TCP communication decrease the maintainability of the software.

It was therefore decided to redesign the data transport software on the HLT to solve the issues and modernize the framework.
To increase the maintainability, core components were built on top of the well maintained, open-source library \zmq, which is an industry-standard for high-performance broker-less asynchronous messaging in distributed applications.
\zmq was already evaluated~\cite{Dworak2011} and applied successfully in several other projects with similar requirements.


All inter- and intra-node communication is now handled by \zmq-based TCP connections.
The buffering is done using the TCP message queue via the \zmq library, so no ring buffers are necessary anymore, removing problems with a residual end-of-run state.
The very general \zmq message format allows us to pass either event data or flow-controlling messages, making it possible to implement features such as back-pressure or fair-queueing.
\zmq handles initialization and cleanup of connections automatically.

\subsubsection{Connections}

The core functionalities of all new components are implemented as generic connections, which are shared among the different parts of the software.
Each connection consists of a pair of generalized sockets.
They are called sender and receiver in the following, although both directions of communication are typically allowed.
Most of the different applications needed for data transport are composed by an input connection of receiver type, which passes the received messages to the application logic, an output connection of sender type, which passes on the result of the application downstream, as well as a connection for external monitoring.
This reusable pattern makes the implementation concise and clean, increases the maintainability and simplifies the addition of new components to the system.
In most cases, the logic is already handled in the shared connection types and the application logic is concise.
All connections as well as the application logic itself run in the same thread.
All multithreading needed for the communication, e.g. for message queuing, is handled by \zmq, simplifying the development of the software.

All messages sent via a connection always consist of three sections, which are internally implemented as three \zmq message parts: a message type, the optional message data and an optional field for additional data.
The message type, consisting of a single byte, defines how the content of the message should be interpreted.
The data section as well as the additional data section are used to transport the actual message, e.g.\ the serialized detector data.
None of the implemented connection types alter nor depend on the content in these sections.
By default, the transported data is uncompressed, as the raw detector data format is already in a quite condensed form.
One exception is the transported DQM information, which is mostly passed in the format of serialized ROOT histograms and has a large potential for compression.
These histograms are compressed using the LZ4 algorithm.

There exist two main types of connections: a load-balanced connection and a confirmed connection.
The load-balanced receiver registers at the sender on start.
For each event it receives from the sender, it sends back a ready message.
By treating all receivers in the order of the received ready messages, the sender transports the event data messages in a load-balanced fan-out fashion.
For buffering, a small amount of ready messages are sent in advance in the beginning.
Control messages such as run start or stop are transported to all registered workers simultaneously.

Senders of a confirmed connection only send the next-event data messages to the receiver once they have received a confirmation message for the current event.
Back pressure produced by the downstream components, e.g.\ the storage system, is therefore passed on upstream.
Confirmed connections can be used for the fan-in part of the data pipeline.

Additional connection types to describe the transportation of JSON-formatted monitoring messages and to connect with external legacy systems via a raw non-\zmq TCP connection are also implemented.

\subsubsection{Data flow}
\begin{figure}
    \centering
    \includegraphics[width=\linewidth]{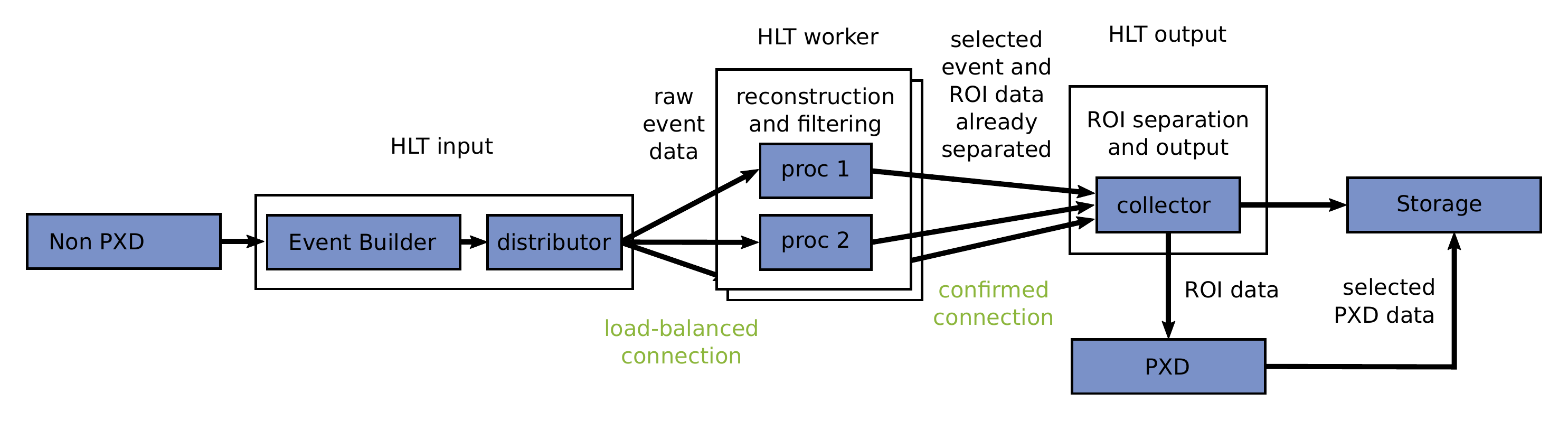}
    \caption{
        Implementation of the HLT data transport using the \zmq library and the implemented connection types.
        Only a single worker with two processes is shown and the DQM histogram transport is omitted for better visibility.
    }
    \label{fig:hlt-zmq}
\end{figure}

Using combinations of the described connections and a minimum of additional logic, the HLT and express reco transport schema can be implemented using \zmq.
\Cref{fig:hlt-zmq} shows the new data flow with the used connection types in the HLT.

On a load request from the global run control to all enabled HLT units, the slow control starts all HLT data transport applications.
After initializing their connections, most of the applications start polling for new messages immediately.
One prominent exception are the basf2 worker processes.
Instead of registering at the distributor, they first initialize all basf2 modules in the reconstruction path, which triggers the loading of the geometry from the conditions database.
After that, as many processes as there are available CPUs on the worker machines are forked out.
Each forked process now registers at the distributor and collector individually.
Only after at least a single process has registered, the run control state is allowed to finish the transition to the ready state.
The data transport applications are now ready to receive the first event without further delay.

Incoming events at the event builder are transported via the distributor to the next ready worker.
The workers decide, after data deserialization and performing the online reconstruction, whether to discard or retain the event.
The event meta data and the raw detector data for retained events are transported with a confirmed connection via the collector to the storage system.
The ROI information extracted during the reconstruction is stored in the same data message in a special format.
It can therefore be transported from the collector to the PXD readout without additional deserialization.

During the reconstruction, defined detector and reconstruction quantities for evaluating the data quality are extracted for DQM.
The histograms are sent as serialized ROOT histograms via confirmed connections.
They are merged in multiple stages and are finally shown in a common DQM server to the shifters and stored for later access.

On run stop, the distributor awaits the delivery of the last event from the event builder and sends stop messages to all workers.
Only if all of those messages reach the collector, the run is stopped successfully and all events are confirmed to be reconstructed on the HLT.

In case of unforeseen problems, the data pipeline can be reset by the shifters.
As the termination of the connections is handled properly by the \zmq library, the shutdown and restart of the applications can be implemented in a straightforward way.
Connection and data flow monitoring give additional information on the stability of the pipeline and introduced timeouts stop the data acquisition automatically in case of unresponsive services.
Additionally, worker processes can be automatically recovered by forking again from the parent process.
As the parent has already preloaded the geometry, the newly created worker process can continue processing events after a delay of maximum \SI{1}{\second}.
The automatic registration of worker processes at distributor and collector is extensible and can also be used in the future for automatic rescaling or rescheduling.

The new implementation was thoroughly tested during the beam downtime in summer 2019.
Long runs above \SI{24}{\hour} and multiple start-stop procedures proved the stability of the system.
As the geometry and condition initialization is now performed before the run start and the histogramprocessing at run stop is optimized, the transition period between runs is only a few seconds instead of minutes.
Beginning in the data taking during winter 2019, the new \zmq implementation will be used on all installed HLT units.

To test the limits of the system, a high rate DAQ test was performed with a single HLT unit.
By removing the event reconstruction and filtering to only test the data transport scheme, rates above \SI{10}{\kilo\hertz} were reached.
This is much higher than the expected \SI{1.5}{\kilo\hertz} per unit in the nominal setup and proves that the \zmq data transport will not cause any problems when increasing the luminosity.
\section{Monitoring and Fault Detection}

For a stable and correct operation of the data acquisition, monitoring and fault detection is implemented in various stages of the data pipeline.
In addition to the aforementioned DQM of physics quantities on the HLT, also technical failures need to be detected.
For this, a multi-stage approach was implemented during the preparation for Belle~II and the first commissioning runs.
Each of the steps -- node monitoring, data flow monitoring and data quality monitoring -- works independently.
The results and alerts of the different stages are closely followed by different expert groups.

\subsection{Node Monitoring}
Hardware properties of important server nodes used during data taking as well as properties of the running operation system and core applications are monitored via a zabbix~\cite{zabbix} installation.
The general-purpose and easily-configurable monitoring framework monitors the machines in a various number of aspects, ranging from temperatures to network properties to the presence of specific necessary running processes on the nodes.
Alerts for errors in the core functionalities, such as the cvmfs~\cite{cvmfs} setup used for software distribution, inform expert shifters and operators directly in case of failing parts.
The alerts are also integrated with the common messaging platform used during data taking operations based on Rocket Chat~\cite{rocketchat}.

The historical data can be used to evaluate the impact of new features after their implementation or can be used to correlate simultaneous problems in different subsystems.
As the setup is easily extendable via user scripts, most of the alerting of the core systems was implemented easily.
Although not in use already, machine learning based predictive maintenance could decrease the downtime of the systems or prevent hardware and software failures.
Tools for the needed time-series analysis exists in many industry applications (e.g. Ref.~\cite{Christ2018}) and can leverage the growing amount of historical data from nearly one year of data taking.

Additional to the quantity-based monitoring of the software and the employed hardware with zabbix, many valuable insights can be gained from the log output produced by the applications.
In the first commissioning phase of the experiment up to now, the log outputs of the applications of the data pipeline were stored in a SQL database.
This database is also used to host the configuration of the data transportation system as well as parts of the run registry.
The log messages were transported from the applications to the database reusing the already present slow control network.
However, due to the growing needs for more detailed logging, this setup leads to a very demanding load on the database servers and the slow control network, which might lead to problems during the operations.
Therefore, a new separate logging collection service is currently under development.
For the implementation, the industry standard elastic stack will be utilized~\cite{elk}.
It includes log transformation and aggregation tooling and can be integrated easily into the current log transportation system.
Additionally, it will be possible to aggregate and collect additional logs from software unrelated to the slow control system.
Dashboards showing the raw logs as well as aggregated performance indicators and the fast search functionality of the lucene database will help expert users in case of problems.
At a later step, the built-in machine learning capabilities of the elastic stack can be used to gain even more insights into the collected log data.
Automated alerting can decrease the time between incidents and a solution or even prevent these problems in the beginning.

\subsection{Data Flow Monitoring}
Besides monitoring a stable operation of the hardware and services in the data taking system, it is important to supervise the applications used for the detector data collection and transportation to prevent instabilities or to recover quickly from unexpected breakdowns.
The NSM~\cite{nsm}- and EPICS~\cite{epics}-based slow control network is utilized for this.
In regular time intervals, the running applications are asked to supply important state variables to the slow control system.
These quantities are both shown to the expert and control room shifters as well as archived with a EPICS archiver for historical and post-mortem analysis.
The stored values are also a documentation of the detector and software configuration during the data taking for correct interpretation of the recorded data.

Beyond this general-purpose monitoring, which is implemented for all important slow control applications, specific monitoring for debugging and improvement purposes exist for many parts of the system.
One example is the described \zmq-based HLT implementation.
As mentioned, all \zmq applications answer with a JSON-encoded dump of their internal state and counters for debugging purposes when sent a request via a specific TCP message.
As the \zmq framework is also available as a python library, simple monitoring and data collection pipelines can be built and executed.
So far, the collected data is only used for manual inspection, but using more sophisticated data processing frameworks, e.g.,\ leveraging the python data processing and machine learning ecosystem, is technically feasible.

\subsection{Data Quality}
The last step in the quality monitoring and fault detection is the data quality monitoring (DQM).
During the reconstruction performed for the HLT decision, important sub-system specific quality monitoring variables and graphs are extracted.
Those quantities are aggregated and evaluated on a dedicated DQM server, where they are accessible to both the control room and subsystem expert shifters. 
User-defined quality assurance logic highlights strange behavior, both in the detector data as well as the reconstruction process. 
This information is used to evaluate the overall quality of the data as well as the specific subsystems.
The historical DQM data can be accessed for reference and studies.
Additionally, a selected subset of event displays is shown to the shifters in the control room for quick visual inspection.

Using the aggregated DQM graphs, the general quality of each subsystem is judged by both the control room shifters and dedicated expert shifters for each run.
This quality information together with general information such as magnetic field values, accelerator settings or detector configurations and conditions are stored in a run registry database.
The stored values are used later during the data reprocessing and skimming - e.g.,\ to produce a list of good runs for physics analyses.
Due to increasing demand for additional quantities in the database, the run registry was rebuilt for the upcoming data taking periods including a user-friendly web view and an additional REST api.

Due to the early stage in the experiment, most of the data quality assurance relies heavily on manual judgments instead of automatic procedures.
With more experience, these processes are expected to change to more automatic evaluation -- perhaps even utilizing machine learning-based approaches in the future.
\section{Summary}

After a successful commissioning phase, Belle~II has recently started its physics data taking period, which will continue for many upcoming years.
A fault-tolerant, monitored and automated data processing pipeline is crucial for a stable and successful data taking.
We employ a real-time reconstruction and data transportation schema for delivering the software-based high level trigger decision.
In this paper, the current and future improvements in the software utilizing the \zmq library as well as in the reconstruction steps needed for the HLT decision are described.
The system is monitored at three levels -- node, data flow and data quality.
Modern, field-tested and independent monitoring applications were or will be implemented to increase the stability and inspectability of the system.
Due to various improvements and the work of many expert groups, the overall setup runs stably and delivers high quality data for future physics analyses.

\section*{Acknowledgements}
We are grateful for the support of the Japan Society for the Promotion of Science (JSPS) and the German Federal Ministry of Education and Research (BMBF).

\bibliographystyle{ws-ijmpa}
\bibliography{online_software}

\end{document}